\documentclass[aps,prb,twocolumn,groupedaddress,showpacs,floatfix,superscriptaddress]{revtex4-1}
\usepackage{epsfig}
\usepackage{amsmath,amssymb}
\usepackage{graphicx}
\usepackage[dvipsnames,usenames]{color}
\usepackage[normalem]{ulem}
\usepackage{soul} 
\emergencystretch=\maxdimen
\begin{document}

\title{Effect of Strain on Charge Density Wave Order in the Holstein
Model}
\author{B. Cohen-Stead}
\affiliation{Department of Physics, University of California, 
Davis, CA 95616,USA}
\email{bwcohenstead@ucdavis.edu}
\author{N.C. Costa} 
\affiliation{International School for Advanced Studies (SISSA),
Via Bonomea 265, 34136, Trieste, Italy}
\affiliation{Instituto de F\'isica, Universidade Federal do Rio de Janeiro
Cx.P. 68.528, 21941-972 Rio de Janeiro RJ, Brazil}
\author{E. Khatami} 
\affiliation{Department of Physics and Astronomy, San Jos\'{e} State University, San Jos\'{e}, CA 95192}
\author{R.T. Scalettar}
\affiliation{Department of Physics, University of California, 
Davis, CA 95616,USA}

\begin{abstract}
We investigate charge ordering in the Holstein model in the presence of
anisotropic hopping, $t_x, t_y=1-\delta, 1 + \delta$, as a model of the
effect of strain on charge density wave (CDW) materials.  Using Quantum
Monte Carlo simulations, we show that the CDW transition temperature is
relatively insensitive to moderate anisotropy $\delta \lesssim 0.3$, but
begins to decrease more rapidly at $\delta \gtrsim 0.4$.  However, the
density correlations, as well as the kinetic energies parallel and
perpendicular to the compressional axis, change significantly for
moderate $\delta$.  Accompanying mean-field theory calculations show a
similar qualitative structure, with the transition temperature
relatively constant at small $\delta$ and a more rapid decrease
for larger strains.  We also obtain the density of states $N(\omega)$,
which provides clear signal of the charge ordering transition at large
strain, where finite size scaling of the charge structure
factor is extremely difficult because of the small value of the order
parameter.
\end{abstract}

\date{\today}

\maketitle

\section{Introduction}

Studies of the effect of strain in charge density wave (CDW) materials
have seen a significant rise in the past several years
\cite{gan16,wei17,gao18}.  The general interest originates from the
ability to tune a strongly correlated insulating phase, inducing
transitions into alternate patterns of charge order, or into metallic
and even superconducting phases. Moreover, by altering the band
structure, the application of strain also provides specific insight into
the nature of a native CDW phase, for instance into the role of Fermi
surface nesting  \cite{johannes06,johannes08}.  Layered transition metal
dichalcogenides (TMDs) are one of the most commonly investigated classes
of CDW materials; their transitions have previously been tuned by
varying the thickness or gate potential
\cite{tsen15,yang14,renteria14,yu15,hollander15,samnakay15}.  In
2$H$-NbSe$_2$ the CDW transition temperature $T_{\rm cdw}$ increases
from $T_{\rm cdw}=33$ K in the bulk to $T_{\rm cdw}=145$ K in a single
layer \cite{xi15}.  A similar, albeit much smaller, effect is seen in
1$T$-TiSe$_2$ \cite{chen15,chen16}.  Strain is therefore useful since it
provides an alternate method for modulating CDW physics.  Indeed,
exploration of the potential use of strain to adjust optical, magnetic
and conductive properties, especially in TMDs, has been referred to as
`strain engineering'.

Much of the existing theoretical work in the area has been within
first-principles density functional theory (DFT).  These studies find
that for 1$T$-TiSe$_2$ the CDW transition temperature can be enhanced
or suppressed with the application of tensile or compressive strain,
respectively \cite{wei17}.  In the latter case, the weakened CDW opens
the door for superconductivity (SC).  This difference in effect is
linked to the distinct behavior of the band gap upon extension versus
compression. For thin layers of TMDs, the intercalation of chemical
compounds between layers, such as Na-intercalated NbSe$_2$, leads to
strain, which has been shown to enhance SC \cite{lian17}.  Initially,
the Na intercalation creates a large electron doping, which contracts
the Fermi surface and causes CDW to disappear.  The subsequent
application of strain increases the density of states at the Fermi
surface and more than doubles the SC transition temperature.

CDW materials, including the TMDs, generally have complex
(\textit{e.g.}\,layered) structures.
The charge ordering may not be commensurate
with the lattice, and may also differ on the surface and within the
bulk.  The application of strain has additional complicating effects,
including changes in the phonon spectrum and of the relative placement
of different orbitals (energy bands).
In particular, 1$T$-VSe$_2$ has a
transition from hexagonal to rectangular charge order with strain, which
seems to originate in the softening of certain phonon modes
\cite{zhang17}.  The aforementioned DFT investigations have explored
many of these details.  

An alternate theoretical approach to DFT which lends complementary
insight into CDW physics is through the solution of simple lattice
Hamiltonians.  One set of models focuses on intersite electron-electron
interactions $V$, as described, for example, by the extended Hubbard
Hamiltonian \cite{sengupta2002,hirsch85,Iglovikov}.  Here, charge order arises
directly from the minimization of the intersite repulsion energy $V$ by
alternating empty and occupied sites.  A more realistic approach for
TMDs, however, would be including electron-phonon interactions, such as
those incorporated in the Holstein \cite{holstein59} or
Su-Schrieffer-Heeger \cite{su79} models.  In these cases, the driving
force for CDW formation is a lowering of the electron kinetic energy
through the opening of a gap in the spectrum.  This energy lowering
competes with the cost in elastic energy associated with phonon displacements.

CDW formation on surfaces and in quasi-2D materials have been motivating
theoretical studies of the Holstein model in two dimensions.  In
addition to the choice of the CDW driving interaction
(electron-electron-like or electron-phonon-like), lattice geometry plays
an important role in the presence of charge ordering.  For instance, for
the Holstein model in a honeycomb lattice, one may show that a finite
critical electron-phonon coupling is required for CDW
\cite{costa18,chen18}, while in the triangular lattice its
ground state exhibits SC \cite{xiang18}.

In view of these simulation results, here we investigate how
charge-charge correlations are affected by deformations
in the lattice, that is, we focus on the effects of strain
on charge ordering.
To this end, we investigate the
Holstein model on a square lattice using determinant quantum Monte Carlo
(DQMC) simulations, and incorporate the most direct effect of strain,
the enhancement of the orbital overlap integral by compression, through
an anisotropy in the hopping in the $x$ and $y$ directions. We find that
although $T_{\rm cdw}$ is relatively insensitive to anisotropy $\delta
\lesssim 0.3$, the density correlations and kinetic energy change
significantly even at small strain. It is only at larger anistropy
$\delta \gtrsim 0.4$ that significant changes in $T_{\rm cdw}$ are
observed. The paper is organized as follows:
in Sec.\,\ref{Section:Model} we present the main features of the
Holstein Hamiltonian, defining the parameters of interest; Sec.\,\ref{Section:MFT} describes and presents results for a mean-field approach,
while DQMC results are presented in Sec.\,\ref{Section:QMC};
in Sec.\,\ref{Section:discussion} we discuss the results and summarize our main conclusions.

\section{The Model}\label{Section:Model}

The Holstein Hamiltonian, which describes electrons interacting locally with ions, is given by
\begin{align} 
\mathcal{\hat H} = 
& -t_x \sum_{\mathbf{i},\sigma} \big(\, \hat d^{\dagger}_{\mathbf{i},\sigma} 
\hat d^{\phantom{\dagger}}_{\mathbf{i}+\hat x, \sigma} + 
\hat d^{\dagger}_{\mathbf{i}+\hat x,\sigma} 
\hat d^{\phantom{\dagger}}_{\mathbf{i}, \sigma} \,  \big)  
\nonumber \\ 
& -t_y \sum_{\mathbf{i},\sigma} \big(\, \hat d^{\dagger}_{\mathbf{i},\sigma} 
\hat d^{\phantom{\dagger}}_{\mathbf{i}+\hat y, \sigma} + 
\hat d^{\dagger}_{\mathbf{i}+\hat y,\sigma} 
\hat d^{\phantom{\dagger}}_{\mathbf{i}, \sigma} \,  \big)  
- \mu \sum_{\mathbf{i}, \sigma} \hat n_{\mathbf{i}, \sigma} 
\nonumber \\ 
& + \frac{1}{2} \sum_{ \mathbf{i} } \hat{P}^{2}_{\mathbf{i}} +
\frac{\omega_{\, 0}^{2}}{2} \sum_{ \mathbf{i} }
\hat{X}^{2}_{\mathbf{i}} + \lambda \sum_{\mathbf{i}, \sigma} \hat
n_{\mathbf{i}, \sigma} \hat{X}_{\mathbf{i}} \,\,.
\label{eq:ham}
\end{align}
Here $\hat d^{\dagger}_{\mathbf{i},\sigma} (\hat d_{\mathbf{i},\sigma}
)$ are creation (destruction) operators for a fermion of spin
$\sigma=\uparrow,\downarrow$ at site $\mathbf{i}$ of a two-dimensional
square lattice.  Thus, the first term represents an electron kinetic
energy (band structure) with hoppings $t_x, t_y$ and dispersion
$\epsilon_{\bf k} = -2 t_x \, {\rm cos}\,k_x -2 t_y \, {\rm cos}\,k_y $.
$\hat P_{\mathbf{i}}$ and $\hat X_{\mathbf{i}}$ describe a local phonon
mode of frequency $\omega_0$ on site $\mathbf{i}$, where the phonon mass
has been normalized to $M=1$.  The electron-phonon coupling $\lambda$, also sometimes reported in terms of $g=\frac{\lambda}{\sqrt{2\omega_0}}$,
connects the electron density $\hat n_{\mathbf{i},\sigma}= \hat d^{\dagger}_{\mathbf{i},\sigma} \hat
d_{\mathbf{i},\sigma}$ for spin $\sigma$ at site $\mathbf{i}$ with the displacement $\hat X_{\mathbf{i}}$, where
$\mu=-\frac{\lambda^2}{\omega_0^2}$ is the chemical potential at
half-filling.

At constant volume, compression along one axis is accompanied by an
expansion in the orthogonal direction.  Thus, in what follows, we set
$t_x=t(1-\delta)$ and $t_y=t(1+\delta)$, a choice which keeps
$t_x+t_y=2t$, and hence the bandwidth $W=4(t_x+t_y)$ constant.  This is
motivated physically by the remarks above, but also allows us to
separate the effect of hopping anisotropy from changes which would
accompany a simple isotropic reduction or enhancement of $W$.

The electron-phonon interaction promotes local pairing of electrons.
This can easily be seen by considering the single site ($t=0$) limit.
Integrating out the phonon degrees of freedom leads to an effective
attraction between the up and down spin fermions $U_{\rm eff} \,
n_{\mathbf{i},\uparrow} n_{\mathbf{i},\downarrow}$, with $U_{\rm eff} =
-\lambda^2/\omega_0^2$.  Associated with this attraction is an
oscillator displacement $\langle X \rangle = -\lambda \langle n
\rangle/\omega_0^2$ where $\langle n \rangle = \langle n_\uparrow +
n_\downarrow \rangle$ is the density.

At strong coupling, local pairs form due to this on-site attraction.
These pairs prefer to organize their placements spatially.  In
particular, as the density approaches half-filling, $\langle n
\rangle=1$, on a bipartite lattice, electron pairs and empty sites
alternate on the two sublattices.  This CDW pattern is favored because
the energy of neighboring occupied and empty sites is lower by
$4t^2/U_{\rm eff}$ relative to two adjacent occupied or empty sites.
This argument closely parallels the one which motivates the appearance
of antiferromagnetic (AF) order in the large $U$ (Heisenberg) limit of
the half-filled repulsive Hubbard model, where well-formed local moments
of up and down spin alternate due to the $J \sim 4t^2/U$ lowering of the
energy relative to parallel spin placement.

There is a further analogy between the Hubbard and Holstein Hamiltonians
at weak coupling.  In the Hubbard model at $U \lesssim W$, AF order is
associated with Fermi surface nesting and a `Slater insulating' phase
-- the opening of an AF gap lowers the electron kinetic energy.
Meanwhile,
for $U \gtrsim W$ one has a Mott insulator in which AF order arises via
$J$.  In the Holstein model, an alternation of phonon displacements
opens a CDW gap, with similar effect.  It is interesting that these
close analogies exist, in the weak coupling limit, despite the fact that
the Holstein Hamiltonian has a second set of (phonon) degrees of freedom
which is absent in the Hubbard Hamiltonian.  Although the Holstein model
has no strong coupling Mott phase, one still expects the CDW ordering
temperature $T_{\rm cdw}$ to decline at large $U_{\rm eff}$ (large
$\lambda$).  This expectation is not realized within the analytic
Eliashberg treatment, but has been observed in quantum Monte Carlo (QMC)
simulations \cite{weber18,zhang19}.

\section{Mean-Field Theory}\label{Section:MFT}

We first solve Eq.\,\eqref{eq:ham} by making an adiabatic approximation
in neglecting the phonon kinetic energy, and then apply a simple
mean-field  {\it ansatz} by letting $\hat X_{\mathbf{i}} \rightarrow x_0
+ (-1)^{\mathbf{i}} \, x_{1}$.  The value $x_0$ describes a
site-independent phonon displacement which is given by
$-\frac{\lambda}{\omega_0^2}$ at half-filling, similar to that described
in the preceding section. Meanwhile $x_1$ is the CDW order parameter: a
nonzero value breaks the symmetry between the two (equivalent)
sublattices.  

Inserting this form into Eq.\,\eqref{eq:ham}, the quadratic Hamiltonian can
be diagonalized.  From the resulting electronic energy levels $E_\alpha$
one can compute the free energy as a function of the order parameter
$x_1$,
\begin{align}
F = \frac{N}{2} \omega_0^2 (x_0^2+x_1^2) - T \sum_{\alpha,\sigma}
{\rm ln} \big( 1 + e^{-\beta E_{\alpha}(x_1)} \big)
\label{eq:F}
\end{align}
Minimizing $F(x_1)$ determines the presence ($x_1>0$) or absence
($x_1=0$) of CDW order.  Since the product of the coupling constant
$\lambda$ and the phonon displacement $x_{\mathbf{i}}$ provides a
staggered chemical potential at site ${\bf i}$, a non-zero value of
$x_1$ will result in an alternating electron density, that is, CDW
order.

An equivalent iterative approach is as follows: given some intial $x_0$
and $x_1$, the (quadratic) Hamiltonian is diagonalized and the resulting
charge densities  $n_{\mathbf{i}} = n + (-1)^{\mathbf{i}}\, \Delta n$
are computed.  Using these values, $x_0$ and $x_1$ are updated via $x_0
= (-\lambda/\omega_0^2) \, n$ and $x_1 = (\lambda/\omega_0^2) \, \Delta
n$.  This process is iterated to convergence.

It is evident that within mean-field theory (MFT)
the behavior of the Holstein model is
governed only by the combination $\lambda^2/\omega_0^2$ rather than on
$\lambda$ and $\omega_0$ individually. This is also the case at $t=0$,
but is only approximately true in exact solutions, \textit{e.g.}\,within DQMC.
Nevertheless, it is convenient to define the dimensionless coupling constant $\lambda_D \equiv
\lambda^2/(\omega_0^2 \, W)$ where $W=8t$ is the fermion bandwidth, and
present results as functions of $\lambda_D$.

\begin{figure}[t]
\includegraphics[scale=0.35]{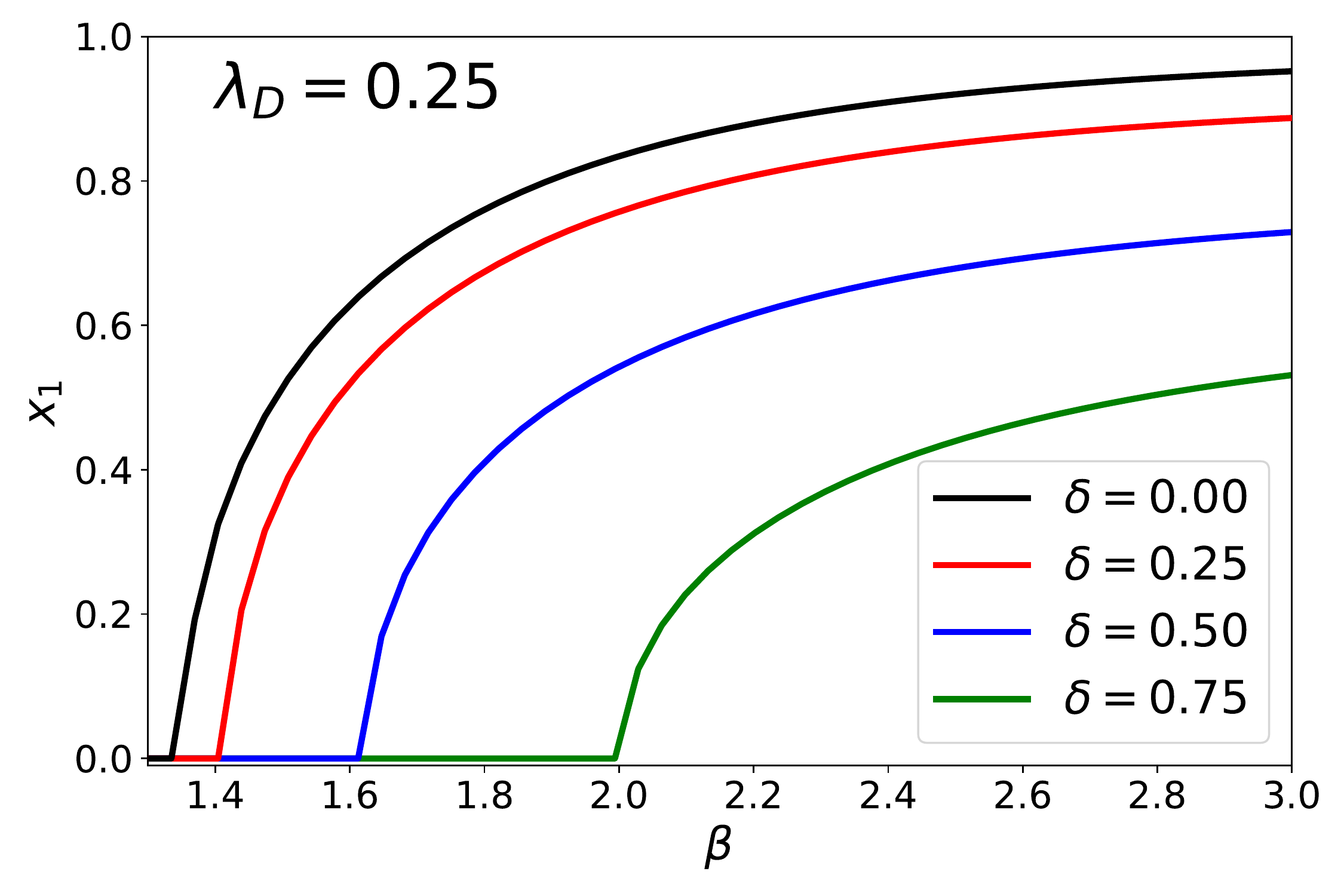} %
\caption{ 
The mean-field order parameter $x_1$ versus $\beta$ for four different
values of strain $\delta$ at $\lambda_D=0.25$.
The critical transition temperature $T_{\rm cdw}=\beta_c^{-1}$
decreases with increasing strain.  See Fig.~\ref{fig:MFT2}.
}
\label{fig:MFT1} 
\end{figure}

\begin{figure}[t]
\includegraphics[scale=0.35]{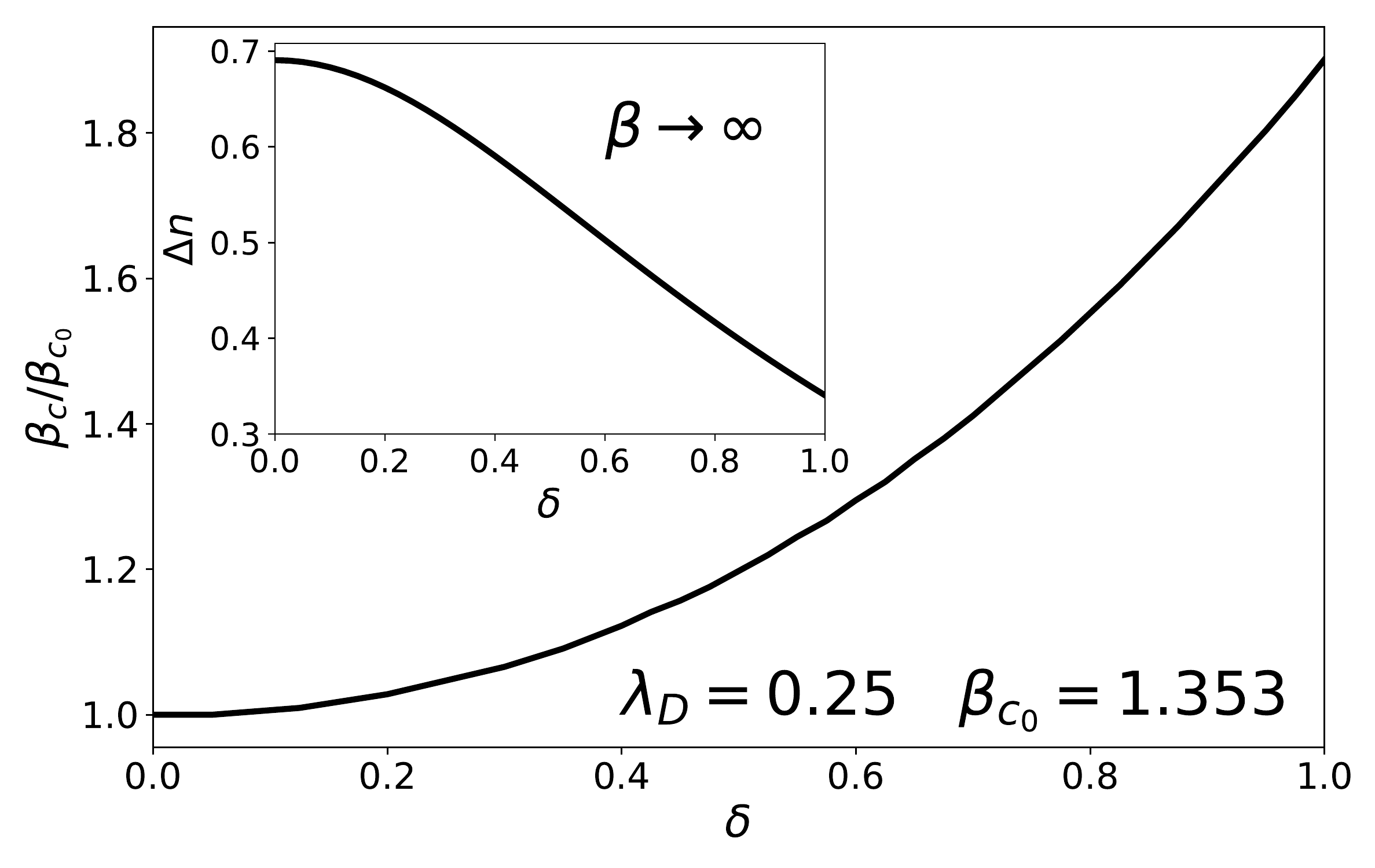} %
\caption{ 
The mean-field critical temperature $\beta_{c}/\beta_{c_0}$ versus the
strain $\delta$ for $\lambda_D=0.25$ where $\beta_{C0}=1.353$ 
in the isotropic $\delta=0$ case. The
inset shows the mean-field result for the difference in electron
density between the two sub-lattices $\Delta n$ in the limit that $\beta
\rightarrow \infty$.  }
\label{fig:MFT2} 
\end{figure}

Figure~\ref{fig:MFT1} shows the MFT behavior of $x_1$ as
function of the inverse temperature for different values of $\delta$,
given lattice size of $L=150$.
Note that, as
expected, there is a finite-temperature second-order phase transition,
and that the maximum value that $x_1$ approaches at low temperatures
changes significantly with $\delta$. This behavior is also reflected in
the inset of Fig.\,\ref{fig:MFT2},
showing that the difference in electron
density between the two sublattices $\Delta n$ decreases with increasing
$\delta$ in the $T \rightarrow 0$ limit.  Because of the $x
\leftrightarrow y$ symmetry, we expect $T_{\rm cdw}(-\delta)=T_{\rm
cdw}(\delta)$, where the change in the critical temperature is a
monotonically decreasing even function of $\delta$.


Since the CDW phase transition in
the Holstein model is at the same
universality class of the 2D Ising model,
it is worth comparing our MFT results (and subsequent DQMC results)
for $\beta_c$ with those from the 2D anisotropic
Ising model, \textit{i.e.\,}$J_x \neq J_y$. 
Within a mean-field approach
for $J_x=1-\delta$ and $J_y=1+\delta$,
one obtains $2 \beta_c (J_x + J_y) = 1$, giving $\beta_c=1/4$ that is completely independent of $\delta$,
in stark contrast to the exact Onsager solution.
Unlike the Ising model, the $\beta_c$ obtained using a mean-field approach for the CDW transition in the Holstein model 
depends on $\delta$.
This occurs because the density of states at the Fermi surface
is modified via the effect of $\delta$ on the band structure.

\section{Quantum Monte Carlo}\label{Section:QMC}
\subsection{Methodology}

We next treat the Hamiltonian of Eq.\,\eqref{eq:ham} with determinant
quantum Monte Carlo method \cite{blankenbecler81,hirsch85,white89a}.  A
detailed discussion of this approach may be found in reviews,
such as Refs.\,\onlinecite{gubernatis16,assaad02,santos03b}.  In
evaluating the partition function ${\cal Z} = {\rm Tr}\,e^{-\beta
\mathcal{\hat H}}$, the inverse temperature is discretized into
$L_{\tau}$ intervals of length $\beta = L_\tau \Delta \tau$.  Complete
sets of phonon position eigenstates $\{ \,|\,x_{\mathbf{i}}(\tau) \,
\rangle \, \}$ are then introduced between each incremental
imaginary-time evolution operator $e^{-\Delta \tau \mathcal{\hat H} }$.
The action of the quantum oscillator pieces in the third line of
Eq.\,\eqref{eq:ham} on $ |\,x_{\mathbf{i}}(\tau) \, \rangle \, $ leads
to the usual ``bosonic' action,
\begin{align}
\nonumber S_{\rm Bose} = & \Delta \tau \bigg[ \, 
\frac{1}{2} \omega_0^2 \sum_{\mathbf{i,\tau}} x_{\mathbf{i}}(\tau)^2 \\
& + \frac{1}{2} \sum_{\mathbf{i,\tau}}  \bigg(
\frac{x_{\mathbf{i}}(\tau+1) - x_{\mathbf{i}}(\tau)}{\Delta \tau} \bigg)^2
\bigg] \, .
\end{align}
The fermionic operators appear only quadratically, and can be
traced out analytically.  The result is the product of the
determinants of two matrices $M_{\sigma}(\{x_{\bf i}(\tau)\})$, 
one for each of spin $\uparrow, \downarrow$.  
The remaining trace over the phonon field involves a sum over the
classical variables $x_{\mathbf{i}}(\tau)$ indexed by the two
spatial and one imaginary-time directions, with a weight
given by 
$e^{-S_{\rm Bose}} \, 
{\rm det} M_\uparrow(\{x_{\bf i}(\tau)\})
{\rm det} M_\downarrow(\{x_{\bf i}(\tau)\})$. This sum is done via a Monte Carlo sampling using both single and global updates.

Because the two spin species couple in the same way to the phonon
coordinates, the matrices $M_\sigma$ are identical for
$\sigma=\uparrow,\downarrow$.  Hence the product of their determinants,
which enters the weight of the configuration $\{ x_{\mathbf{i}}(\tau)
\}$, is always positive, ensuring there is no `sign problem'
\cite{loh90,troyer05} at any temperature, density or Hamiltonian
parameter values.  Nevertheless, in order to emphasize the effects of
strain, we limit our analysis to the half-filling case, \textit{i.e.}\,$\langle
n_{\mathbf{i}\sigma} \rangle=\frac{1}{2}$, where a commensurate CDW phase is known to
exist below a given critical temperature \cite{weber18}.

The principle limitations of DQMC, as with most Monte Carlo simulations,
are finite lattice sizes and statistical error bars on the observables.
One way in which finite size errors manifest in DQMC
is via the discrete set of momentum points $\{ {\bf k} \}$.  Here we use
antiperiodic boundary conditions for lattices with linear size $L=6$, 10
and 14 and periodic boundary conditions for $L=4$, 8 and 12.  This
ensures that the four ${\bf k}$ points
$(\pm\frac{\pi}{2},\pm\frac{\pi}{2})$ fall directly on the Fermi Surface
for all lattice sizes, mitigating otherwise substantial finite size
effects.
\begin{figure}[t]
\includegraphics[scale=0.34]{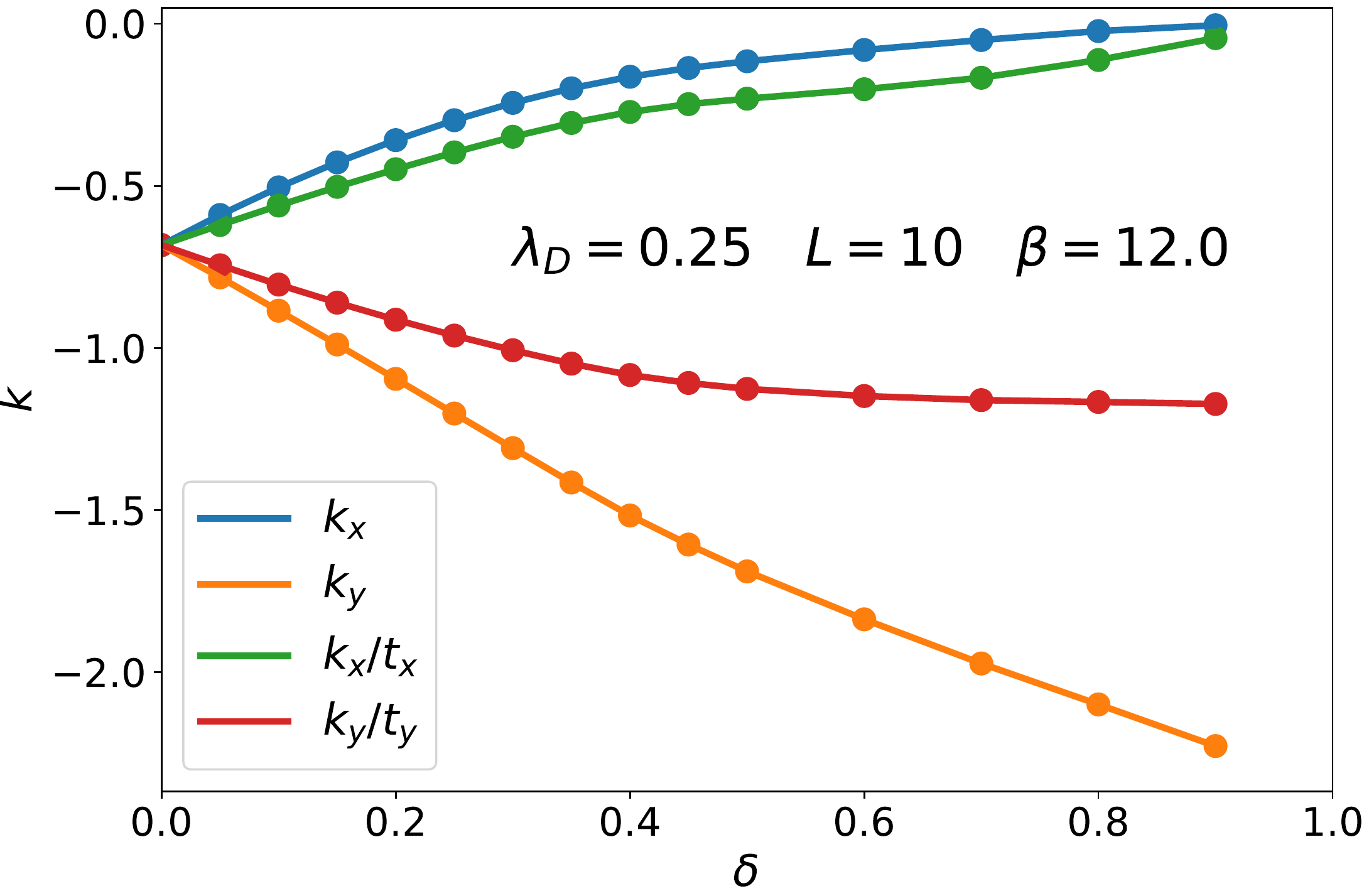} %
\caption{ 
The electron kinetic energies $k_x$ and $k_y$ are shown as functions of
$\delta$.  Division by the energy scales $t_x$ and $t_y$ isolates the
effect of anisotropy on the hopping.
}
\label{fig:DQMC_KE} 
\end{figure}

\begin{figure}[t]
\includegraphics[scale=0.34]{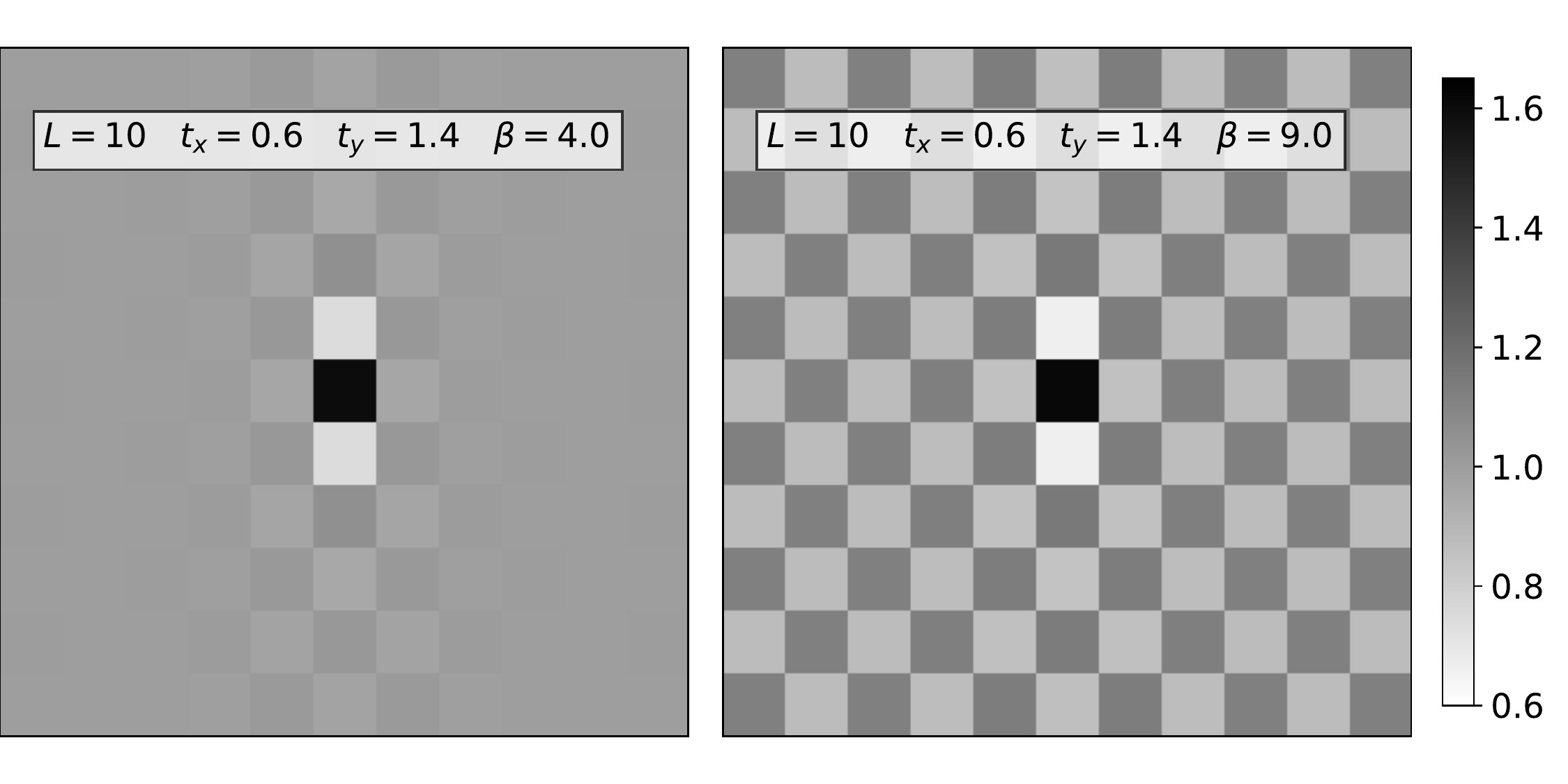} %
\caption{
\underbar{Left Panel}:
Real space density-density correlations for a moderate strain of
$\delta=0.4$ at $T>T_{\rm cdw}$. Note the enhanced correlations in the
$\hat y$ direction relative to the $\hat x$ direction.  \underbar{Right
Panel}: Real space density-density correlations for
$\delta=0.4$ at $T<T_{\rm cdw}$. Note that the oscillating checkerboard
charge density pattern now persists across the entire lattice.
}
\label{fig:DQMC_CR1} 
\end{figure}

\begin{figure}[t]
\includegraphics[scale=0.4]{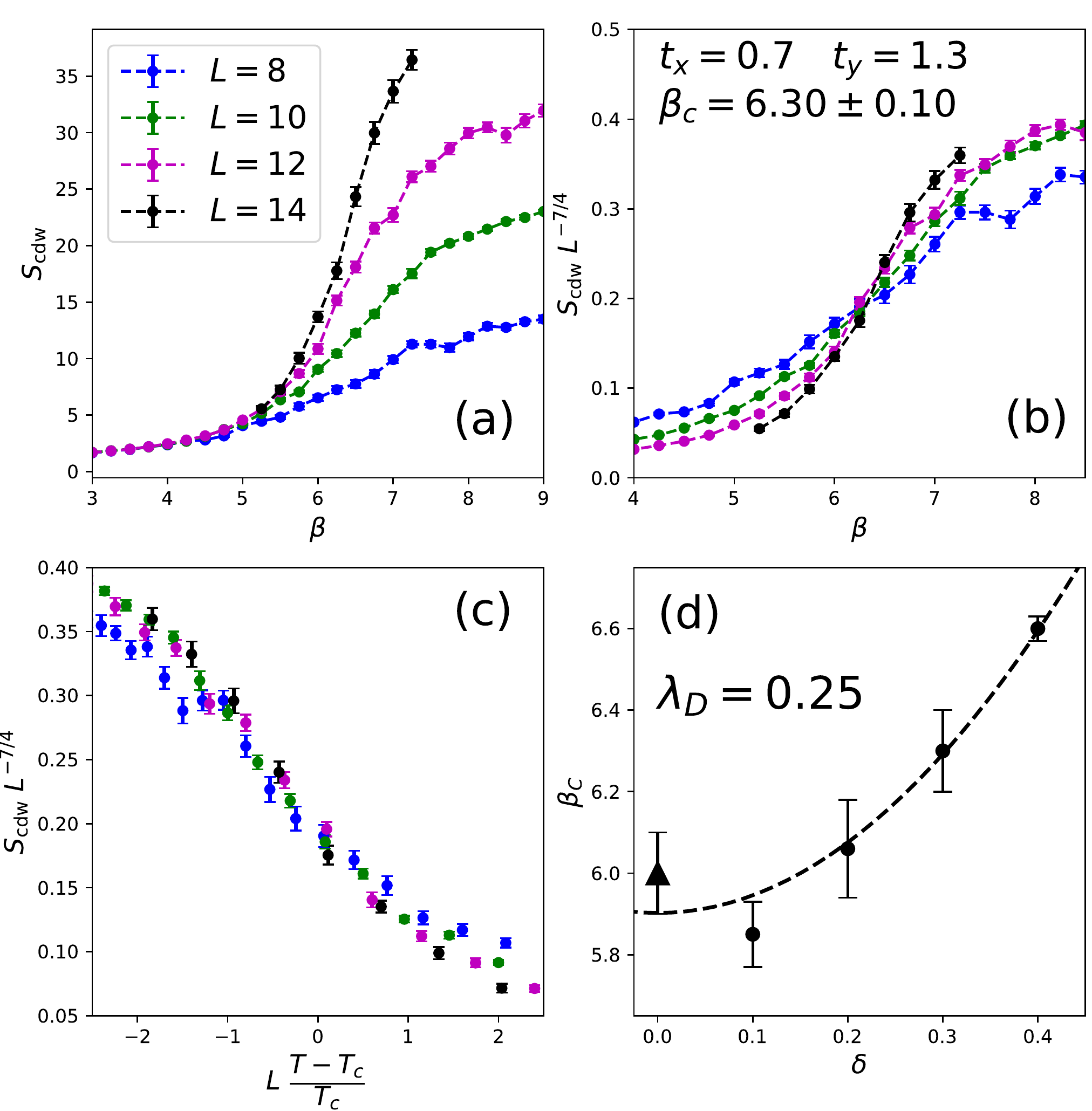} %
\caption{
\underbar{Panel (a):} $S_{\rm cdw}$ versus $\beta$ for $\delta=0.3$, $\lambda_D=0.25$ and four different lattice size.
\underbar{Panel (b):} A finite size scaling where the scaled structure factors $S_{\rm cdw}L^{-\gamma/\nu}$ 
exhibit a crossing as a function of $\beta$ for
different lattice sizes $L$. We infer $\beta_c = 6.3 \pm 0.1$ is
slightly increased from the isotropic $\beta_c = 6.0$.
\underbar{Panel (c):} The full data collapse in which the temperature axis is also
scaled by $L^{1/\nu}\left( \frac{T-T_{\rm cdw}}{T_{\rm cdw}} \right)$.
\underbar{Panel (d):} $\beta_c$ as a function of $\delta$. The dashed
line is a least squares fit to the data.  The value of 
$\beta_c$ at $\delta=0$ (triangle) is from Ref.~\onlinecite{costa18}.
}
\label{fig:DQMC_FSS} 
\end{figure}

Using DQMC, we are able to access a wide variety of observables, since
expectation values of fermionic operators are straightforwardly expressed
in terms of matrix elements of $G_\sigma = M_\sigma^{-1}$ and their products.
In what follows, we consider first the kinetic energies in the $x$ and $y$
directions,
\begin{align}
k_x &\equiv  \big\langle \, -t_x  \sum_\sigma \big( \,
\hat d^{\dagger}_{\mathbf{i},\sigma} 
\hat d^{\phantom{\dagger}}_{\mathbf{i}+\hat x,\sigma} 
+ \hat d^{\dagger}_{\mathbf{i}+\hat x,\sigma} 
\hat d^{\phantom{\dagger}}_{\mathbf{i},\sigma} 
\, \big)
\, \big\rangle
\nonumber \\
k_y &\equiv  \big\langle \, -t_y  \sum_\sigma \big( \,
\hat d^{\dagger}_{\mathbf{i},\sigma} 
\hat d^{\phantom{\dagger}}_{\mathbf{i}+\hat y,\sigma} 
+\hat d^{\dagger}_{\mathbf{i}+\hat y,\sigma} 
\hat d^{\phantom{\dagger}}_{\mathbf{i},\sigma} 
\, \big)
\, \big\rangle 
\label{eq:ke}
\end{align}
and the staggered CDW structure factor
\begin{align}
S_{\rm cdw} &= \frac{1}{N} \sum_{\mathbf{i,r}} (-1)^{\bf r}
\big\langle \big( \, n_{\mathbf{i}\uparrow} +n_{\mathbf{i}\downarrow} \, \big)
\,
\big( \, n_{\mathbf{i+r}\uparrow} +n_{\mathbf{i+r}\downarrow} \, \big)
\big\rangle \nonumber \\
&= \frac{1}{N} \sum_{\mathbf{i,r}} (-1)^{\bf r} c(\mathbf{r}) \, ,
\label{eq:S}
\end{align}
which is the Fourier transform at $\mathbf{q}=(\pi,\pi)$ of the real
space density correlation functions $c(\mathbf{r})$, and is proportional
to the square of the order parameter when extrapolated to the
thermodynamic limit. When making these measurements we use $\Delta \tau = 0.125$, which is small enough that the Trotter errors associated with the discretization of $\beta$ are smaller than the statistical ones.\cite{Scalettar91}

\subsection{Equal-Time Correlations}

The kinetic energy directly measures the effect of strain via an
anisotropic hopping in the $x$ and $y$ directions.  We will also display
$k_x/t_x$ and $k_y/t_y$ to isolate the `trivial' factor of the energy
scales.  Figure \ref{fig:DQMC_KE} shows the kinetic energies as
functions of the hopping anisotropy $\delta$.  These evolve smoothly
with $\delta$, increasing in the $y$ direction, for which $t_y =
1+\delta$, and decreasing in the $x$ direction, where $t_x = 1-\delta$.

The real space density correlations $c(\mathbf{r})$ are given in
Fig.~\ref{fig:DQMC_CR1} for a $10 \times 10$ lattice at temperatures
both above and below $T_{\rm cdw}$ for anisotropy $\delta=0.4$. For
$T<T_{\rm cdw}$  the correlations extend over the entire lattice in a
checkerboard pattern expected for $(\pi,\pi)$ ordering. However, in the
$T>T_{\rm cdw}$ case the correlations extend further in the $y$
direction than the $x$ direction, indicating that charge ordering
forms first in the direction of enhanced hopping.

\begin{figure}[t]
\includegraphics[scale=0.34]{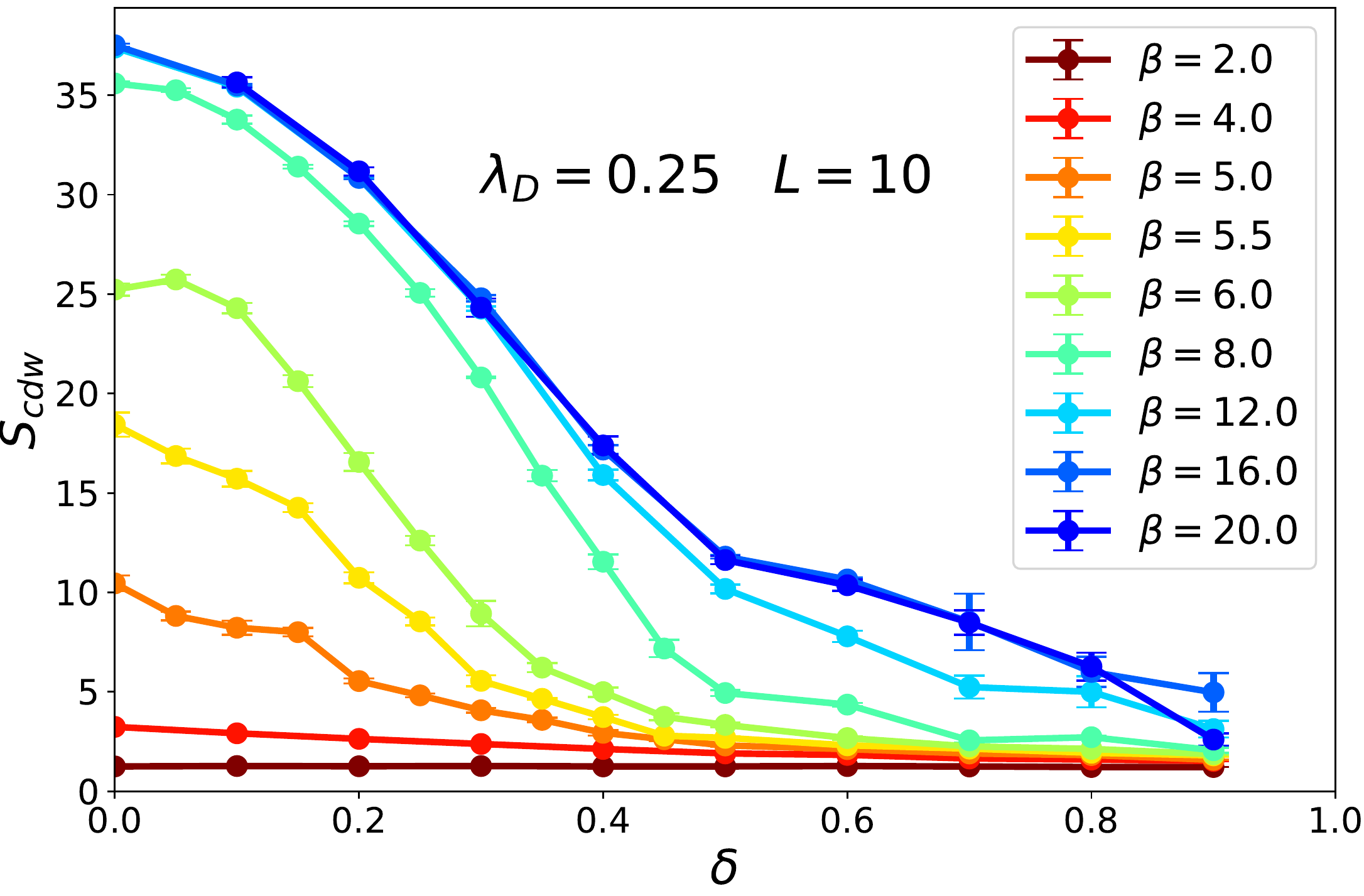} %
\caption{
CDW structure factor versus hopping anisotropy $\delta$.  
The low temperature value of the CDW order parameter falls
to approximately half of its isotropic value as $\delta
\rightarrow 0.4$.
}
\label{fig:DQMC_SCDW} 
\end{figure}

The CDW structure factor $S_{\rm cdw}$  is sensitive to the development
of long-range change order. At high temperature,
density correlation $c(\mathbf{r})$ in the disordered phase is short ranged, and $S_{\rm cdw}$ is of order unity. On the other hand, in
the CDW phase, density correlations extend over the entire lattice and
$S_{\rm cdw} \sim N$.
This change in behavior is illustrated in
Fig.~\ref{fig:DQMC_SCDW} for different values of $\delta$.
For the isotropic case ($\delta = 0$) it occurs at an energy scale
$\beta \sim 6/t$, but as $\delta$ increases, the onset of CDW
order is deferred to lower temperatures.

A finite size scaling of $S_{\rm cdw}$ allows a more precise
identification of $T_{\rm cdw}$.  This task is considerably simplified
by the knowledge that the appropriate universality class is that of the
2D Ising model, since CDW order breaks a two-fold discrete symmetry on
the square lattice \cite{weber18,zhang19,chen18}.
Results are shown for $\delta=0.3$ in Fig.~\ref{fig:DQMC_FSS}\,(b). $\beta_c$ is inferred from the crossing of $L^{-7/4}S_{\rm cdw}$ for
different linear lattice sizes $L$, and Fig.~\ref{fig:DQMC_FSS}\,(c) shows the associated collapse of the of the $S_{\rm cdw}$ data.  Fig.~\ref{fig:DQMC_FSS}\,(d) gives
$\beta_c $ for the range $0.0 < \delta \lesssim 0.4$.
For $\delta=0.0$, $\beta_c$ is taken from Ref.
\onlinecite{costa18}, which is consistent with more recent simulations
using the Langevin method to evolve the phonon fields \cite{batrouni18}.
$\beta_c$ for all $\delta>0.0$ was obtained by the associated crossing
plots. However, as $\delta$ increases we find finite size effects
increase and, as a consequence, smaller lattice sizes could no longer
be used in the crossing; the ranges of lattice sizes used to extract the
critical temperature for each $\delta$ are shown in the table below.

\begin{center}
 \begin{tabular}{|c||c|c|c|c|} 
 \hline
 $\delta$ & 0.1 & 0.2 & 0.3 & 0.4 \\
 \hline
 $L_{\rm min}$ & 6 & 8 & 8 & 10 \\
 \hline
 $L_{\rm max}$ & 12 & 12 & 14 & 14 \\
 \hline
\end{tabular}
\end{center}

One might naively expect that $T_{\rm cdw}$ would scale as $t^2/U_{\rm eff}$,
the energy scale which reflects the difference between a doubly occupied and
empty site being adjacent relative to two doubly occupied or two empty
sites.  The kinetic energy measurement of Fig.~\ref{fig:DQMC_KE}
gives a sense of how this quantity varies in the $x$ direction.
At $\delta=0.5$ it is lower by a factor of roughly three,
so that $T_{\rm cdw}$ might be expected to be reduced by an order of magnitude
from $T_{\rm cdw} \sim t/6$ in the isotropic case.  However, this almost certainly
underestimates $T_{\rm cdw}$ as it ignores the enhancement 
of density correlations in the $y$ direction.  Nevertheless these
estimates seem consistent with Fig.~\ref{fig:DQMC_SCDW}, which shows that it is challenging to detect CDW order $\delta \gtrsim 0.5$, even at temperatures as low as $\beta t=24$, four times the isotropic $\beta_c$.

The small structure factors shown in Fig.~\ref{fig:DQMC_SCDW} for large strain, even at low temperatures, reflect a significant increase in $\beta_c$ as $\delta \rightarrow 1$. For $\beta \, t=20$, $S_{\rm cdw}$ is
less than $1/20$ of its value for perfect classical charge order.  Some initial insight into this is given by
the MFT results, where as $\beta \rightarrow \infty$ the greatly reduced
value of $S_{\rm cdw}$ at large $\delta$ is reflected in the smallness
of the MFT order parameter $x_1$.
In the next section, we will present data suggesting that the behavior
of $N(\omega)$ provides more definitive evidence of the persistence of
the CDW insulating phase even at large strain.

\subsection{Spectral Function}

The spectral function can be obtained from the Green's function
measurement in DQMC combined with analytic continuation \cite{jarrell96}
to invert the integral relation
\begin{align}
G({\bf k},\tau) = \int d\omega \frac{A({\bf k},\omega) \ e^{-\tau \omega} }
{e^{-\beta \omega} + 1}
\label{eq:apw}
\end{align}
Following the procedure discussed in Ref.\,\onlinecite{white91}, one can evaluate
the moments
\begin{align}
\mu_1 ({\bf k}) &\equiv \int d\omega \, \omega \, A({\bf k},\omega) 
 \\ \nonumber
&=
\big( \, \epsilon_{\bf k} - \mu \,) + \lambda \, \langle X \rangle
 \\
\label{eq:muapw}
\mu_2 ({\bf k})  &\equiv \int d\omega \, \omega^2 \, A({\bf k},\omega) 
 \\ \nonumber
&=
(\epsilon_{\mathbf{k}}-\mu)^2+2\lambda(\epsilon_{\mathbf{k}}-\mu)\langle X \rangle + \lambda^2 \langle X^2 \rangle
\end{align}
Here $\langle X \rangle$ is the phonon displacement on a spatial site,
and is related to the density by $\langle X \rangle = -\lambda \langle n
\rangle / \omega_0^2$.  At half-filling, $\langle n \rangle=1$ and $\mu
= U_{\rm eff} = -\lambda^2/\omega_0^2$ so that $\mu_1({\bf k}) =
\epsilon_{\bf k}$.  (This is the same as for the noninteracting case,
since there $A({\bf k},\omega) = \delta(\omega-\epsilon_{\bf k})$.)
These analytic values of the moments, in combination with a measurement
of the phonon potential energy, serve as a useful check on the analytic
continuation. Preliminary tests indicate analytic continuation of the imaginary-time dependent Greens function
obtained from DQMC yields values for the moments in agreement with the
analytic results of Eq.~(\ref{eq:muapw}) to within a few percent.

\begin{figure}[t]
\includegraphics[scale=0.34]{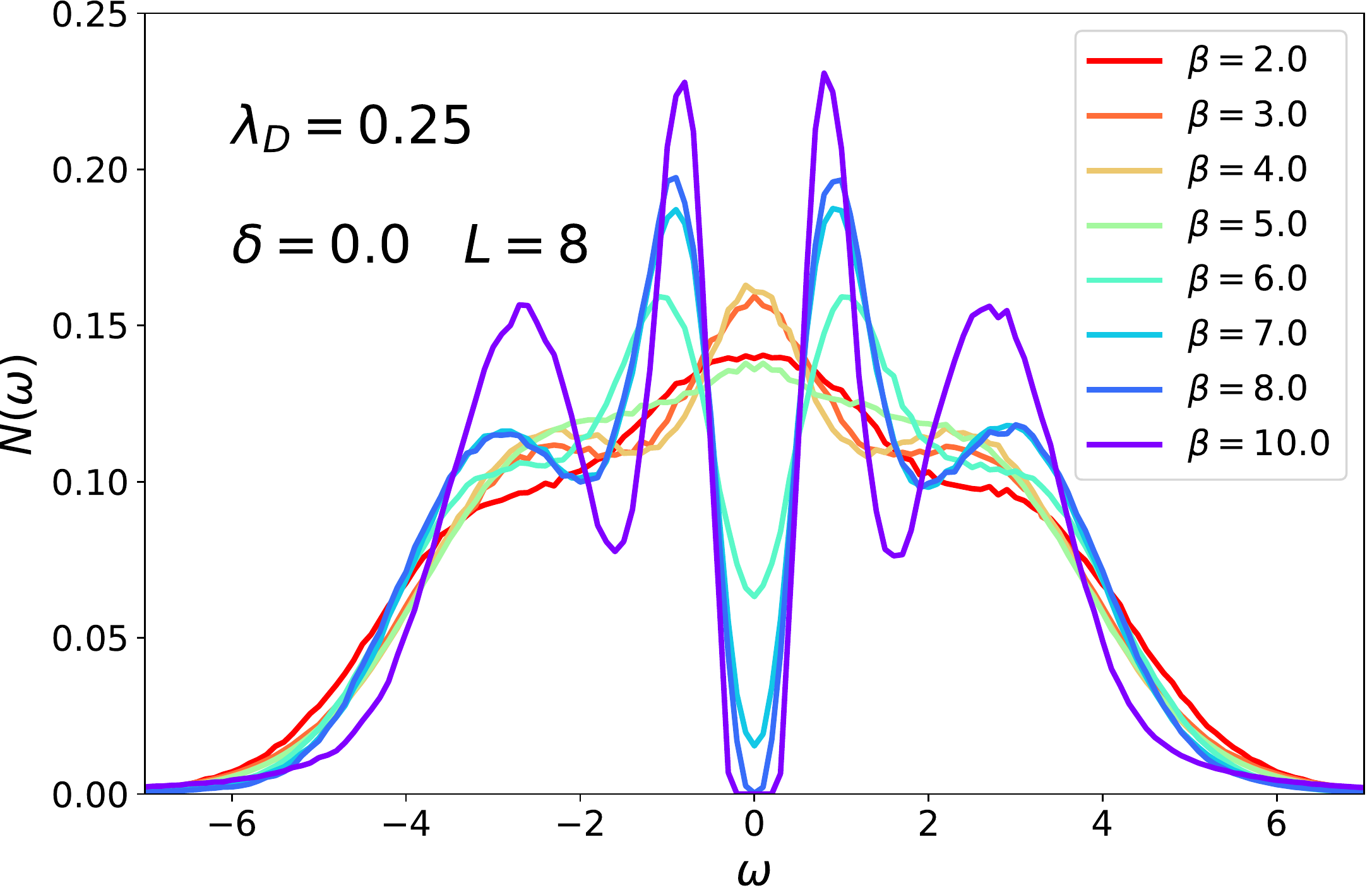} %
\caption{
Density of states for the isotropic lattice for different
inverse temperatures $\beta t$.
The phonon frequency $\omega_0=t$ and electron-phonon
coupling $g= \, t$.  Finite size scaling of $S_{\rm cdw}$ suggests
$\beta_c \, t= 6.0 \pm 0.1\,\,$  \cite{costa18}, which is consistent
with the $\beta$ value at which a full gap opens in $N(\omega)$.
}
\label{fig:DQMC_DOSdelta0} 
\end{figure}

\begin{figure}[t]
\includegraphics[scale=0.34]{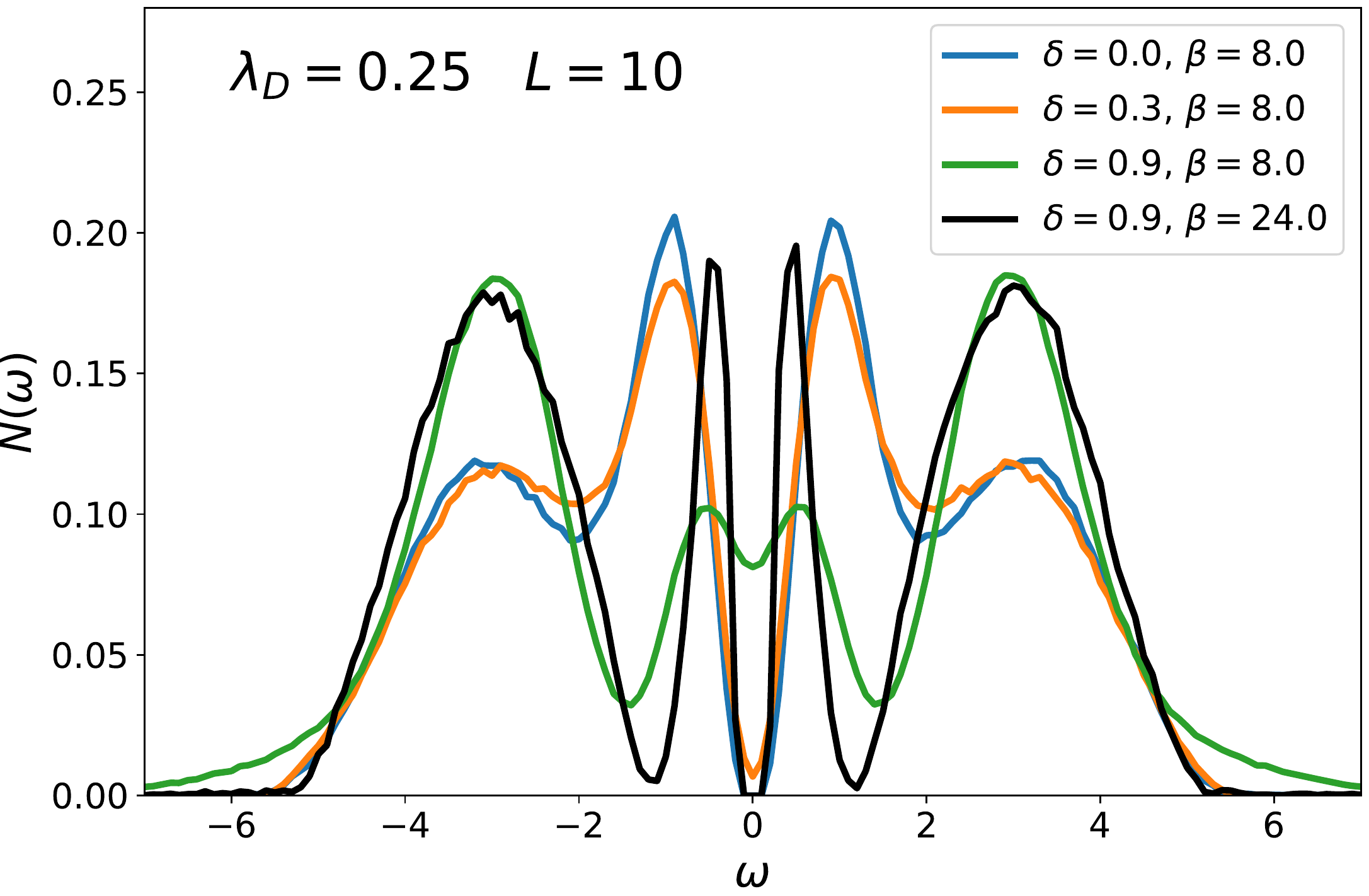} %
\caption{
Density of states comparing the isotropic lattice with
small ($\delta=0.3$) and
large ($\delta=0.9$) anisotropy.
For $\delta=0.9$ the opening of a gap is delayed until
$\beta_c \, t\sim 20$.
}
\label{fig:DQMC_DOSdeltaneq0} 
\end{figure}

Figure \ref{fig:DQMC_DOSdelta0} shows the density of states $N(\omega)$
for the isotropic lattice.  At inverse temperatures $\beta \, t=2,3,4,5$
(\textit{i.e.}\,lower than $\beta_c t$), $N(\omega)$ has a peak at the Fermi level
$\omega =0$.  Beginning at the critical inverse temperature inferred
from the finite size scaling of $S_{\rm cdw}$ \cite{costa18},
$N(\omega)$ develops a gap, which provides another indication of the
transition to the insulating CDW phase.
Fig.~\ref{fig:DQMC_DOSdeltaneq0} shows that $N(\omega)$ remains
relatively unchanged under the influence of strain $\delta=0.3$,
consistent with the robust $S_{\rm cdw}$ of Fig.~\ref{fig:DQMC_SCDW} at
modest anisotropy.  However, at $\delta=0.9$ the CDW gap has been
replaced by a weak minimum at $\beta \, t=8$ and is only recovered at
$\beta \, t =24$.

The formation of a gap at $\delta=0.9$, even though the corresponding
$S_{\rm cdw}$ value shown in Fig.~\ref{fig:DQMC_SCDW} is small, is
strong evidence that a CDW insulating phase persists out to very large
$\delta$. It is useful to consider the two-dimensional Ising model when
trying to understand this result. The Onsager solution gives a non-zero
$T_{c}$ for all $J_x/J_y>0$ in the Ising model, a result consistent
with the general expectation that anisotropy in the form of a weak
coupling in one direction does not  destroy a finite temperature second
order phase transition in dimension $d$.  The rough physical picture is
that correlations will develop in the `strongly interacting' directions
out to a length $\xi$.  The coordinated orientation of degrees of
freedom in regions of size $\xi^{d-1}$ then creates a large `effective'
coupling $J_{\rm eff} \sim \xi^{d-1} J_{\rm small}$ in the weakly
interacting direction.  As $\xi$ grows, $J_{\rm eff}$ eventually boosts
$J_{\rm small}$. This same argument can be applied to the CDW order in
the Holstein model, a claim supported by Fig.~\ref{fig:DQMC_CR1}
showing that for $T>T_{\rm cdw}$ density correlations first form in the
direction of enhanced hopping.

\section{Discussion}\label{Section:discussion}

In this work we investigated charge ordering in the Holstein model on a
square lattice in the presence of anisotropic hopping, $t_x,
t_y=1-\delta, 1 + \delta$. For $\delta \lesssim 0.3$, the transition
temperature $T_{\rm cdw}$ remains relatively stable, only decreasing
significantly for $\delta \gtrsim 0.4$. However, both the electron
kinetic energies and the structure factor $S_{\rm cdw}$ see significant
shifts for small values of $\delta$. The suppression of $S_{\rm cdw}$,
especially at larger strains, mirrors the smallness of the MFT order
parameter $x_1$ with increasing $\delta$. Despite the smallness
of $S_{\rm cdw}$ at low temperatures and large $\delta$, the opening of
a gap in the density of states $N(\omega)$ at $\delta=0.90$ indicates
the presence of an insulating CDW transition even as $\delta \rightarrow
1$.

While we have focused here exclusively on the effects of anisotropic
electron hopping $t_x \neq t_y$ on charge correlations and the gap in
the Holstein model, it is also possible to examine the role of changes
in the phonon spectra.   Indeed, DFT calculations \cite{wei17} indicate
that such changes, \textit{e.g.\,}enhancement of the phonon frequency with
compression, are central to the onset of CDW order.  Similarly, it is
known from DQMC simulations that $T_{\rm cdw}$ exhibits a non-monotonic
dependence on $\lambda_D = \lambda^2/(\omega_0^2 \, W)$ in the Holstein
Hamiltonian \cite{zhang19}.  The possibility of direct connection of
such model calculations to materials would require the introduction of
a connection of $\omega_0$ (and $\lambda$) to strain.

Applications of DQMC to Hamiltonians with repulsive electron-electron
interactions are limited by the sign problem \cite{loh90,troyer05};
study of Holstein or Su-Schrieffer-Heeger models with electron-phonon
interactions are much less restricted.  As seen here, and in other work
\cite{weber18,zhang19,chen18}, low enough temperatures can be reached to
get a complete understanding of the CDW transition, and even of the
possibility of quantum critical points \cite{zhang19,chen18} associated
with CDW transitions driven by changes in $\lambda_D$ at $T=0$.  Recent
work has further exhibited this flexibility of DQMC by examining the
effects of phonon dispersion on CDW order in the Holstein model
\cite{costa18}.  In short, the freedom from the sign problem opens the
door to incorporating additional materials details into quantum
simulations of electron-phonon models and hence to the study of CDW
transitions.  Such rich details are much more difficult to include in
studies of repulsive electron-electron interactions like the Hubbard
model for which the sign problem is severe.

The density of states $N(\omega)$ gives information about the CDW gap.
However, the momentum-resolved spectral function $A({\bf k},\omega)$
yields more detailed data concerning the effect of (strain) hopping
anisotropy on the quasiparticle dispersion, and in particular, the
possibility that gaps might develop at distinct temperatures as the momentum ${\bf k}$ changes. Work to study that possibility is in progress.

\vskip0.03in \noindent
\underbar{Acknowledgements:}  The work of B.C-S. and R.T.S.  was
supported by the Department of Energy under grant DE-SC0014671.
N.C.C. was partially supported by the Brazilian funding agencies
CAPES and CNPq.
E.K. acknowledges support from the
National Science Foundation (NSF) under Grant No. MR-1609560.
Computations were performed in part on Spartan high-performance
computing facility at San Jos\'{e} State University, which is
supported by the NSF under Grant No. OAC-1626645.


\bibliography{references}

\end{document}